# Comparative Analysis of the Electronic Energy Structure of Nanocrystalline Polymorphs of $Y_2O_3$ Thin Layers: Theory and Experiments

D.W. Boukhvalov[1,2*], D.A. Zatsepin[2,3], Yu. A. Kuznetsova[2], N.V. Gavrilov[4], A.F. Zatsepin[2]

[1] *College of Science, Institute of Materials Physics and Chemistry, Nanjing Forestry University, Nanjing 210037, P. R. China*
[2] *Institute of Physics and Technology, Ural Federal University, Ekaterinburg, Russia*
[3] *Institute of Metal Physics, Ural Branch of Russian Academy of Sciences, Ekaterinburg, Russia*
[4] *Institute of Electrophysics, Ural Branch of Russian Academy of Science, Ekaterinburg, Russia*

The results of fabrication and characterization of atomic structure of nanocrystalline thin layers of $Y_2O_3$ in cubic and monoclinic phases is reported. Experimental data demonstrate crystalline ordering in nanocrystalline films with average grain size of ~10-14 nm both for cubic and monoclinic studied structures. Density Functional Theory (DFT) based simulations demonstrate insignificant differences of electronic structure of these phases in the bulk and on the surfaces. Theoretical modeling also pointed out the significant broadening of valence and conductive bands caused by means of energy levels splitting in agreement with experimental data (X-ray photoelectron and photoluminescence spectra). The presence of various intrinsic and extrinsic defects (including surface adsorption of carbon mono- and dioxide) does not promote visible changes in electronic structure of $Y_2O_3$ surface for both studied phases. Optical absorption and luminescence measurements indicate insignificant bandgap reduction of $Y_2O_3$ nanocrystalline layers and the very little contribution from defect states. Simulation of extrinsic compression and expanding demonstrate stability of the electronic structure of nanocrystalline $Y_2O_3$ even under significant strain. Results of comprehensive studies demonstrate that yttrium oxide based nanocrystalline layers are prospective for various optical applications as a stable material.

E-mail: danil@njfu.edu.cn

## 1. Introduction

Yttrium oxide thin-films have wide application importance since they usually exhibit good chemical and thermal stability with high melting point, fracture toughness, high refractive index, relatively wide band-gap (see, i.e., Refs. [1-4]), etc. These features allow to employ $Y_2O_3$ films as technologically suitable coatings for thermo-protective barriers, electric gate interfaces and others for different commercial applications [5-6]. In addition, $Y_2O_3$ is used as a host material for doping with different ions, especially with rare-earth and transition ions [7, 8]. Optical properties of $Y_2O_3$-based materials usually well meet manufacturing requirements for creation of light-emitting diodes [9], color displays [10], lasers [11], solar cells [12], bio-imaging systems [13], etc.

It is known that yttrium oxide can exist in three polymorphic forms denoted as A-type (hexagonal), B-type (monoclinic) and C-type (cubic) structures [14]. Most of published papers deal with the study of cubic $Y_2O_3$ properties, because it is stable at room temperature and standard atmospheric pressure. Also relatively simple synthesis conditions are required in order to yield cubic polymorph what is one more technological advantage. Monoclinic and hexagonal yttrium oxides have been studied to a much lesser extent if compared with cubic polymorph. This circumstance is usually associated with technological difficulties of A- and B-type $Y_2O_3$ synthesis: here the need to apply elevated temperatures (up to 2600 K) and high pressures (up to 32 GPa) [15, 16] exists. However, even these extreme synthesis conditions cannot guarantee the yield of single-phase final material. Due to technological difficulties of a single-phase monoclinic $Y_2O_3$ synthesis, the properties of this polymorph had been poorly studied if compared with that for cubic $Y_2O_3$. Additional pitfall for growing $Y_2O_3$ thin-films on a substrate is fairly strong dependence of growing film on size effects in the vertical axis, which increases the difficulty in optimization of mechanical properties [17]. For this reason, strict grade control of final product with surface sensitive research methods is needed.

The features of electronic and optical properties of any material essentially depend on atomic and energy structure arrangements: crystal symmetry type, defectiveness, electronic and phonon density of states. For example, it was shown in Ref. [18] that monoclinic $Y_2O_3$ films are characterized by slightly larger band gap and slightly lower refractive index if compared with that for cubic $Y_2O_3$ films. The influence of point defects in the structure of cubic $Y_2O_3$ on luminescent characteristics has been studied in Refs. [19, 20]. It was found that both oxygen vacancies and interstitial oxygen in the cubic $Y_2O_3$-host usually lead to a decrease in luminescence quantum yield of europium [19] and bismuth [20] doping ions. At the same time, we have to note that there is no data concerning the influence of defectiveness on optical properties of monoclinic $Y_2O_3$ polymorph, what can be considered as a separate challenging scientific task. In the case of monoclinic yttrium oxide, this issue is of particular importance, since the fabrication of stable monoclinic phase is possible only when non-stoichiometry of oxygen sublattice exists, as was shown in Ref. [21].

In the current paper we consider issues related to the comparative study of energy structure of cubic and monoclinic Y$_2$O$_3$ films, as well as the influence of various types of defects on the formation of optical properties of these Y$_2$O$_3$ polymorphs. We will apply experimental methods of X-ray Photoelectron Spectroscopy and Optical Spectroscopy in combination with DFT-based simulations for data analysis. These analyzed data will provide necessary groundwork for onward studies of thin-film materials based on yttrium oxide with different crystal structures.

## 2. Samples and Experimental Details

Y$_2$O$_3$ films were deposited on silica glass substrates using dc-pulsed mode (50 kHz, 10 μs) of reactive magnetron sputtering technology. Prior deposition, silica glass substrates were cleaned in ultrasonic bath employing acetone solution for 20 min onward drying in airflow. In order to obtain a film with cubic structure, a target with a diameter of 40 mm and a thickness of 2 mm was prepared by cold pressing of yttrium metal powder at 30 MPa. As for monoclinic film, sodium was added during preparation of the target. The purpose of adding sodium is to stabilize monoclinic Y$_2$O$_3$ phase. Usually, an addition of alkali ions into rare-earth oxide is employed to create non-stoichiometry in oxygen sublattice [22]. As it was reasonably shown in Ref. [21], oxygen vacancies promote the nucleation of monoclinic Y$_2$O$_3$ phase. Sodium with an expected ~5 at. % concentration was introduced into the coating at the stage of its deposition. To obtain sodium vapor, a soda-lime glass plate was heated to temperatures close to the glass transition temperature. Plasma sodium vapor content was monitored by the amplitude of the sodium doublet line λ=589.0 nm in the plasma emission spectrum.

The magnetron, sputtered target, and the substrate were placed in a vacuum chamber pumped down to 6.6×10$^{-3}$ Pa with the help of turbomolecular pump. Sputtering was carried out for 8 hours at magnetron power of 30 W in argon-oxygen atmosphere with 0.4 Pa total pressure (the volume concentration of oxygen was less than 30 %). The temperature of substrate was 400 ± 25°C during deposition procedure. After deposition has been completed, the sample was cooled down to room temperature in vacuum chamber at 10$^{-5}$ Torr pressure. The thicknesses of Y$_2$O$_3$ films were determined to be ~800 nm employing ball-abrasion method with the help of Calotest device (CSM Instruments SA, Switzerland).

Structural-phase analysis of the samples under study was performed by means of X-ray diffraction using XPertPro MPD diffractometer with Cu Kα = 1.5405 Å radiation. Diffraction patterns obtained (see Fig. 1) confirm the formation of single-phase cubic Y$_2$O$_3$ film (lattice parameter $a$ = 10.73 Å) [23] and single-phase monoclinic Y$_2$O$_3$ film (lattice parameters: $a$ = 14.07 Å, $b$ = 3.32 Å and $c$ = 8.68 Å) [24]. The processing of diffraction pattern was performed using TOPAS 3 full-profile analysis program accounting the presence of predominant orientation of crystallites (texture) in the material under study. Being parallel to surface the planes (111) and (201) were found as preferred orientation for cubic and monoclinic Y$_2$O$_3$, respectively.

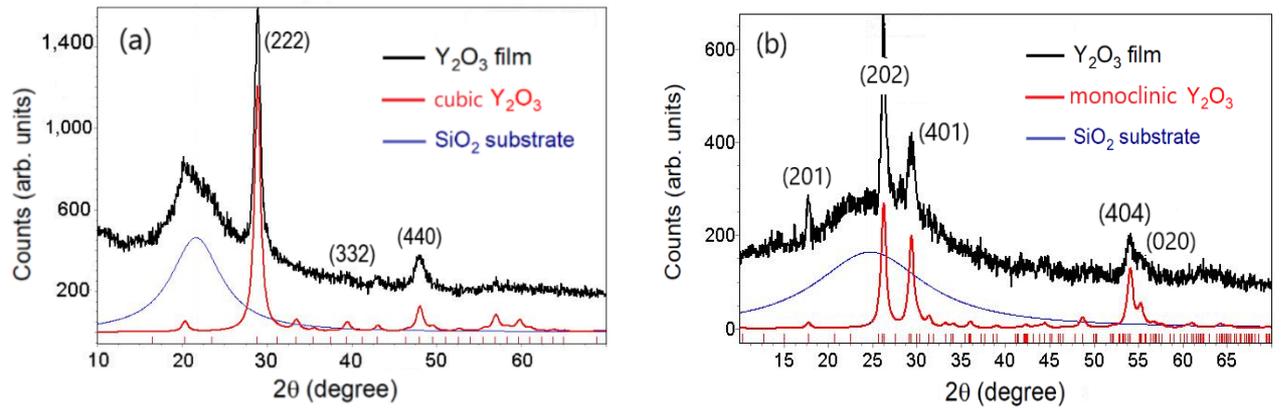

**Figure 1.** X-ray diffraction patterns of cubic
(a) and monoclinic (b) Y$_2$O$_3$ films. Reflexes were identified in accordance with JSPDS No. 43-1036 (for cubic Y$_2$O$_3$ phase) and
JSPDS No. 44-0399 (for monoclinic Y$_2$O$_3$ phase) Standard Cards. A wide diffuse peak originates from silica glass substrate.

**Table I.** Structural-phase characteristics of studied Y$_2$O$_3$ films.

| Film | Crystal structure | Lattice Parameters | Average size of coherent scattering |
|---|---|---|---|
| Y$_2$O$_3$ | cubic, Ia3 | $a$ = 10.73 Å | 10 ± 2 nm |
| Na:Y$_2$O$_3$ | monoclinic, C/2m | $a$ = 14.07 Å, $b$ = 3.32 Å, $c$ = 8.68 Å | 13 ± 2 nm |

The average sizes of coherent scattering $D$ were determined to be 10 nm and 13 nm for cubic and monoclinic films, correspondingly. Assuming that nanocrystallites have a spherical shape, we can estimate surface-to-volume ratio as 0.30 ± 0.05 nm$^{-1}$ and 0.23 ± 0.05 nm$^{-1}$ for nanocrystallites of cubic and monoclinic Y$_2$O$_3$ films, respectively. Table I summarizes structural-phase characteristics of the samples under study.

Since yttrium oxide films are widely used in different optical applications (see Introduction section), it is logically to apply surface-sensitive methods in order to inspect chemical composition and grade of the surface of our samples. For this reason, we employed X-ray Photoelectron Spectroscopy (XPS) using survey and core-level analysis with the help of ThermoScientific *K*-alpha Plus XPS spectrometer. This spectrometer has monochromatic microfocused Al *K*α X-ray source and has 0.05 at. % element sensitivity [25]. An operating pressure in analytic chamber during measurements was not worse than $1.1 \times 10^{-6}$ Pa. Dual-channel automatic charge compensator (GB Patent 2411763) was applied to exclude the charging of our sample under XPS analysis because of the loss of photoelectrons. Pre-run up procedures performed included standard degassing of the sample and analyzer binding energy scale inspection and re-calibration (if needed) employing sputter cleaned Au ($4f_{7/2}$ band), Ag ($3d_{5/2}$ band), and Cu ($2p_{3/2}$ band) inbuilt XPS Reference Standards according to ISO 16.243 XPS International Standard and XPS ASTM E2108-00 Standard. We used ThermoScientific XPS spectrometer inbuilt electronic database, NIST XPS Standard Reference Database [26] and Handbook of Monochromatic XPS Spectra: The Elements of Native Oxides [27] to identify precisely survey spectrum structure (see Fig. 2).

It is well seen from Figure 1 that most intensive peaks belong to yttrium and oxygen which are the components of yttrium oxide. At the same time Na (5.51 at.%) is present monoclinic yttrium oxide film what is the feature of the synthesis technology employed as it has been reported above. As for carbon contamination, it is present in both samples due to well-known ability of $Y_2O_3$ to absorb CO and $CO_2$ species from atmosphere (see, i.e., Ref. [28] and general comments of ThermoScientific on yttrium oxide XPS study [29]). The concentrations of carbon according to our measurements are different for monoclinic and cubic polymorphs: they are 5.56 and 3.97 at.%, respectively. The measured O/Y ratio gives the values 1.46 and 1.50 for monoclinic and cubic $Y_2O_3$, respectively. These values are in a good agreement with data for monoclinic and cubic $Y_2O_3$ reported in Refs. [30-31]. We have to note that freshly synthesized samples are inspected now and all these values obtained well coincide with our previous findings reported earlier in Refs. [23-24]. This means good technological reproducibility of monoclinic and cubic $Y_2O_3$ films with the help of synthesis employed.

**Figure 2.** X-ray photoelectron survey spectra of monoclinic and cubic $Y_2O_3$ films.

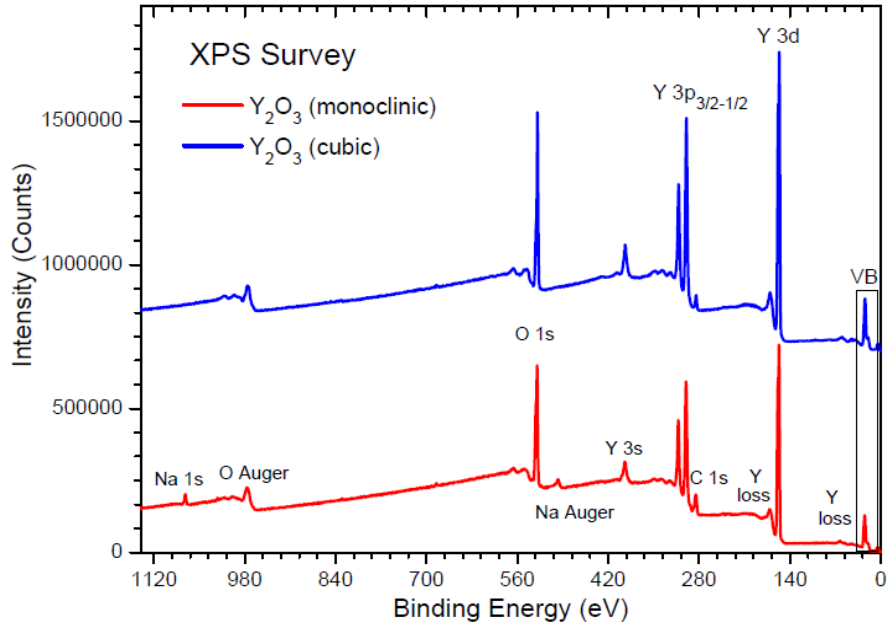

Theoretical modeling of pristine, defective and oxidized $Y_2O_3$ was performed with the use of QUANTUM-ESPRESSO pseudopotential code [32] employing the generalized gradient approximation [33] for exchange-correlation potential in a spin-polarized mode. A full optimization of atomic positions was carried out. During optimization, the electronic ground state was consistently found using ultra-soft pseudopotentials [34]. In order to simulate adsorption of CO and $CO_2$ species by yttrium oxide, an additional correction for London dispersion forces (+vdW) were used. [35] The forces and total energies were optimized with accuracies of 0.04 eV Å$^{-1}$ and 1.0 meV/supercell, respectively. For the modeling of bulk, we used 3×3×3 supercell of cubic or monoclinic $Y_2O_3$ phase (see Fig. 3). For the modeling of surface we used the slabs constructed from the same supercell.

Calculations of formation energies of defects and adsorption were performed using standard formula:

$$E_{form} = (E(products) - [E(host) \pm E(impurities)])/m, \qquad (1)$$

where E(products) is the total energy with defects or adsorbed molecules; E(host) is total energy of the system before formation of defects or adsorption of molecules, and E(impurities) denotes total energies of adsorbed $m$ molecules or formation of $m$ defects in lattice. In the case of interstitial impurities incorporation the sign of total energies of defects in mentioned above formula is plus and in the case of vacancies formation the sign is minus. In the case of multiple defects E(impurities) is the superposition of energies of removed and/or inserted defect atoms. For adsorption of carbon monoxide E(impurity) denotes total energy of single molecule in gaseous phase. The energy value employed for calculations of oxygen defects is total energy of oxygen molecule in gaseous phase divided by two. The energy of sodium impurity was calculated by the following formula:

$$E(Na) = (E(Na_2O) - E(O_2)/2), \qquad (2)$$

where $E(Na_2O)$ denotes total energy of bulk $Na_2O$. Note that negative values of formation energies correspond to exothermic processes whereas positive values correspond to endothermic processes.

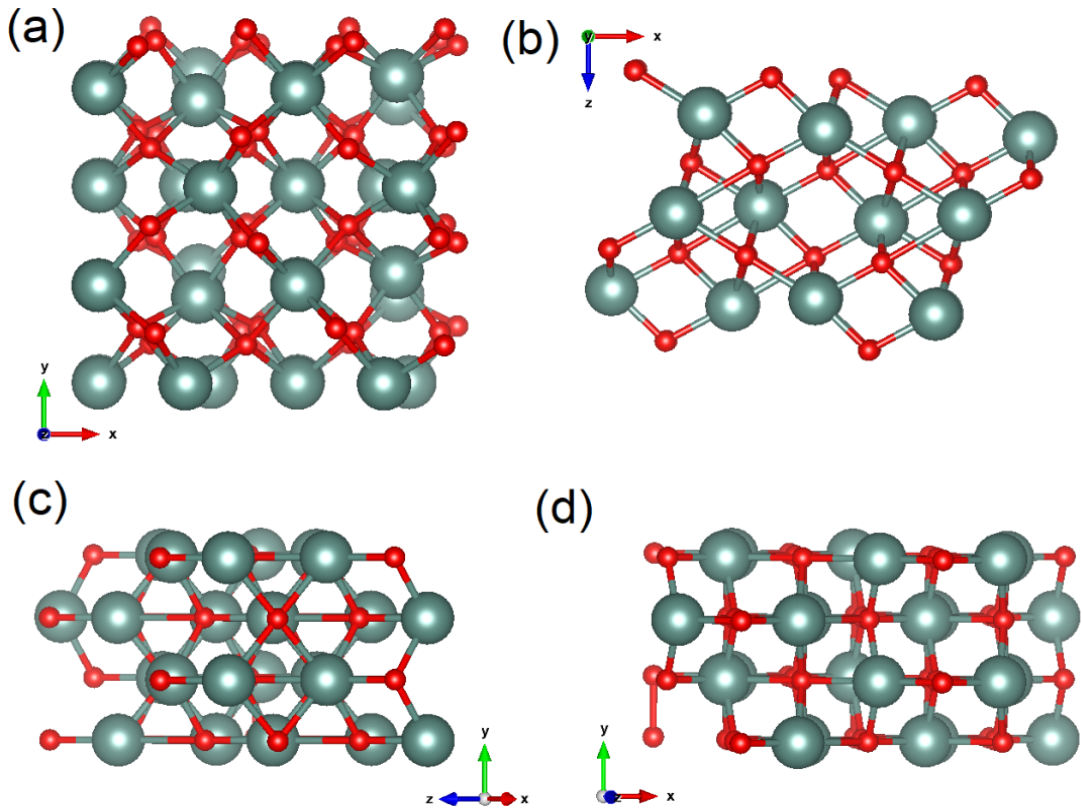

**Figure 3.** Top view of optimized atomic structure of cubic (a) and different views optimized atomic structures monoclinic (b-d) $Y_2O_3$ polymorphs. Yttrium atoms are denoted by means of large grey-blue spheres whereas oxygens are shown as small red spheres.

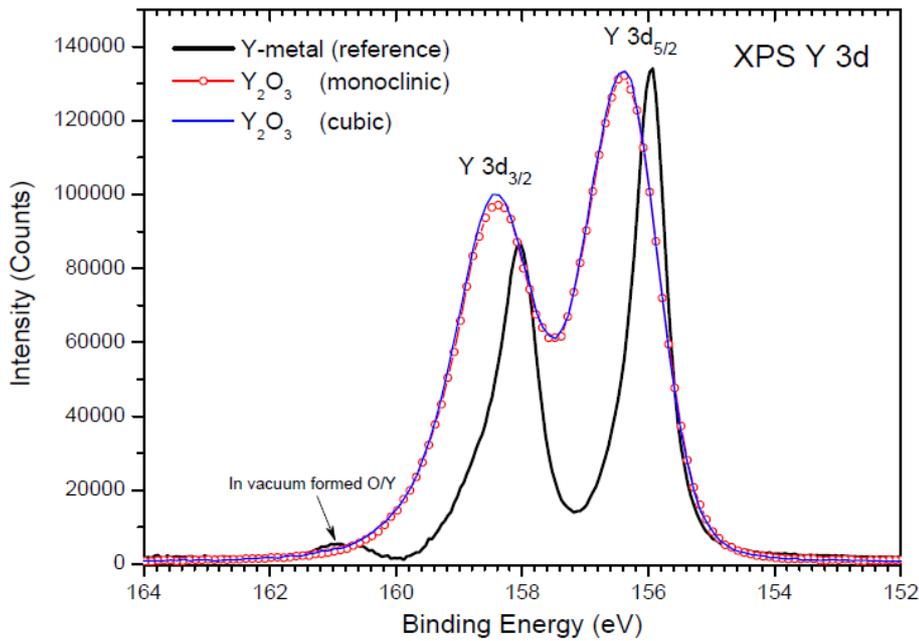

**Figure 4.** X-ray Photoelectron Y 3d core-level spectra of monoclinic and cubic $Y_2O_3$ films. Also XPS Y-metal XPS External Standard Y3d spectrum is shown for comparance.

### 3. Results and Discussion

### 3.1. XPS spectral analysis

X-ray Photoelectron (XPS) Y 3d core-level spectra of monoclinic and cubic $Y_2O_3$ films are shown in Fig. 4. These spectra have well-resolved spin-orbit 5/2 and 3/2 which are located at 156.27 and 158.25 eV, respectively. Appropriate Y 3d core-level of pure Y-metal of XPS External Standard after ThermoScientific are located at more visible lower binding energies (BE = 155.98 eV and 157.99 eV) what is allowing to state that no separate Y-metal phase is present in our samples under study, i.e. samples were synthesized without even partial segregation of Y-metal. This conclusion well coincide with XRD data reported earlier that also indicate single-phase yield of monoclinic and cubic polymorphs of $Y_2O_3$ films due to the synthesis technology employed [23-24]. At the same time well seen that XPS could not detect the differences in Y 3d core-levels between monoclinic and cubic films of $Y_2O_3$ films which is somewhat unusual at first glance. Actually cubic and monoclinic polymorphs of yttrium oxide have quite different lattice parameters ($a = b = c = 10.73$ Å [23] and $a = 14.07$ Å, $b = 3.32$ Å and $c = 8.68$ Å [24]) and crystal structures, respectively, but valence state of yttrium is the same in these polymorphs and large number of Y–O bonds have almost the same values (approximately 2.2 Å according to calculated atomic structures, see Fig. 2 and details of calculation procedure) so this might be the reason for above mentioned situation. Generally speaking, the same story occurs with XPS of rutile and anatase polymorphs of TiO2 where as well not possible to distinguish Ti 2p core-levels among each other (see, e.g., Fig.3a in Ref. [36]) despite visible significant differences in atomic structure but again very similar values of interatomic distances and angles. Additionally, surface and subsurface layers of both $Y_2O_3$ polymorphs could be modified by absorbed CO and $CO_2$ species [28-29] so the dissimilarities between them are leveled by surface defects and newly formed bonds C–O–Y what is impeding XPS analysis of spectra differences. We will inspect all mentioned above statements using O 1s and C 1s core-level spectra analysis and, of course, the onward theoretical modeling with the help of DFT-calculations of electronic densities of states and atomic structure simulations of bulk and surface.

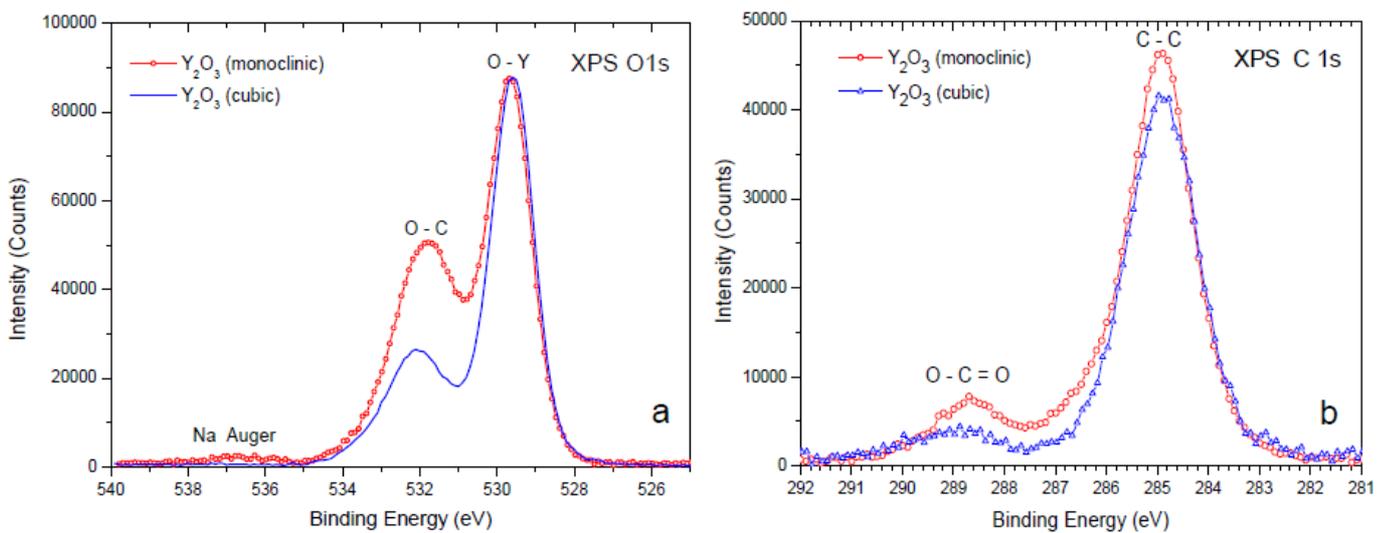

**Figure 5.** X-ray photoelectron core-level O 1s (a) and C 1s (b) spectra of monoclinic and cubic polymorphs of $Y_2O_3$ films.

XPS O 1s and C 1s core-level spectra are shown in Fig. 5 a-b. XPS O 1s spectra exhibit double-band structure with the main band located at 529.86 eV and sub-band located at 532.12 eV (see Fig. 5a). These BE-values well coincide with data reported by P. Yu et. al. [21] and NIST XPS Standard Reference Database [26]. Weak and broad Na Auger band is present in O 1s core-level spectrum due to embedded sodium in this sample, because sodium was used in order to promote the nucleation of monoclinic $Y_2O_3$ phase (see Ref. [21, 24]) and sample preparation details in the current paper). The most remarkable fact is that 532 eV sub-band has dissimilar intensities in O 1s core-level spectra of monoclinic and cubic polymorphs of $Y_2O_3$ films. Since this sub-band is linked with oxygen-carbon imperfections of oxygen sublattice [21] due to absorbed CO and $CO_2$ species, an appropriate "oxygen-carbon" sub-band should be present in C 1s core-level spectra of monoclinic and cubic polymorphs of $Y_2O_3$ films. Exactly this sub-band one can easily see at 288.68 eV in Fig. 4b which is linked with O–C=O bonds [26, 37]. Additionally, exactly monoclinic polymorph O 1s and C 1s spectra have higher intensity of O–C and O–C=O bands, respectively, because of different O/Y relation for monoclinic and cubic $Y_2O_3$ and differently absorbed carbon (see Samples and Experimental Details section and Refs. 22-23). This means a good agreement among our core-level spectra and data of outside researchers cited above.

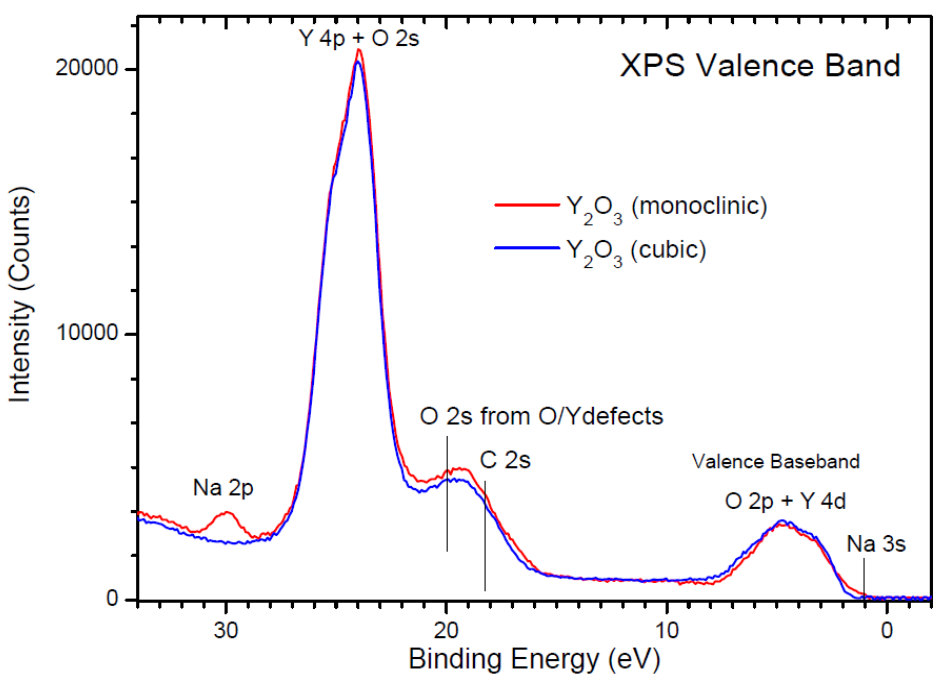

**Figure 6.** X-ray photoelectron Valence Band (VB) spectra of monoclinic and cubic polymorphs of $Y_2O_3$ films.

At the final stage of our XPS analysis we performed Valence Band (VB) mapping of electronic states for monoclinic and cubic polymorphs of $Y_2O_3$ films (Fig. 6). This mapping was made using crossed-referencing among NIST XPS Standard Reference Database [26], Handbook of Monochromatic XPS Spectra [27], ThermoScientific XPS Knowledge Base [37], ThermoScientific XPS spectrometer inbuilt electronic database and Atomic Calculation of Photoionization Cross-Sections and Asymmetry Parameters for Al Kα X-ray probe (recall, we have used Al Kα X-ray Microfocused source for excitation of our x-ray photoelectron spectra) [38]. The most important now, to establish the majority of electronic states in strongly overlapped areas of Valence Bands. Overlapped Y 4p + O 2s states area of binding energies is characterized by the majority contribution of Y 4p electronic states because of the following photoionization cross-sections relation: σ(Y 4p) : σ(O 2s) = 1.2 : 0.19 [38]. The O 2s + C 2s area in the range from 21.5 eV up to 16 eV exhibit σ(O 2s) : σ(C 2s) = 0.19 : 0.066 cross-sections relation. This means that the contribution of carbon C 2s into VB of studied $Y_2O_3$ films is very weak in this BE-region and sub-band located at 19.4 eV (see Fig. 6) is predominantly of O 2s character [38]. The most interesting is mapping of Valence Baseband region. Here relation is σ(Y 4d) : σ(O 2p) : σ(Na 3s) = 4.5 : 2.4 : 0.0097 [38] so formally Valence Baseband is related with Y 4d majority. Finally, valence bands were identified as the bands with Y 4p and Y 4d majority contributions. We suppose that weak differences in VB's are exactly due for this reason, because yttrium is in the same valence state in both polymorphs of studied $Y_2O_3$ films. Of course, all these interpretations of VB mapping will be inspected by means of electronic densities of states calculations in theoretical part of current paper.

## 3.2. Theoretical modeling

An inspection of suitability of used approach for the modeling of yttrium oxide was selected as the first step of our simulations. For this purpose we performed calculations of bulk $Y_2O_3$ which has cubic and monoclinic structures. The deviations between calculated and experimentally measured values of lattice vectors are below 1%. Calculated total energy per formula unit of monoclinic phase is 70.1 meV and this value is higher than that for cubic one. Thus calculations performed also suggest smaller stability of monoclinic yttrium oxide.

Next step of our modeling is the calculation of formation energies of various impurities in the bulk morphology of both phases of $Y_2O_3$ – cubic and monoclinic. We also considered formation of oxygen defects in interstitial void positions ($O_i$, see Fig. 7) additionally to conventional defects in oxides such as oxygen vacancies ($v_O$). Since stabilization of monoclinic phase requires some Na-doping so the formation of substitution ($Na_Y$) and interstitial ($Na_i$) defects was also simulated. For the sake of comprehensiveness all possible combinations of sodium and oxygen defects were simulated in both phases. Results of the calculations performed (see Table II) demonstrate that formation of defects in cubic phase of $Y_2O_3$ is extremely unfavorable and requires energies above 5.7 eV per defect. Thus we can conditionally consider cubic phase as stable and almost defect-free. Contrary, the formation of defects in less stable monoclinic Y$_2$O$_3$ requires smaller values of energy. For some kind of defects (such as interstitial oxygen and pair of interstitial oxygen and sodium defects, see Fig. 7) formation energies are relatively moderate: they are 1.08 and 2.04 eV/defect, respectively. Therefore, an appearance of defects mentioned in monoclinic Y$_2$O$_3$ is very likely.

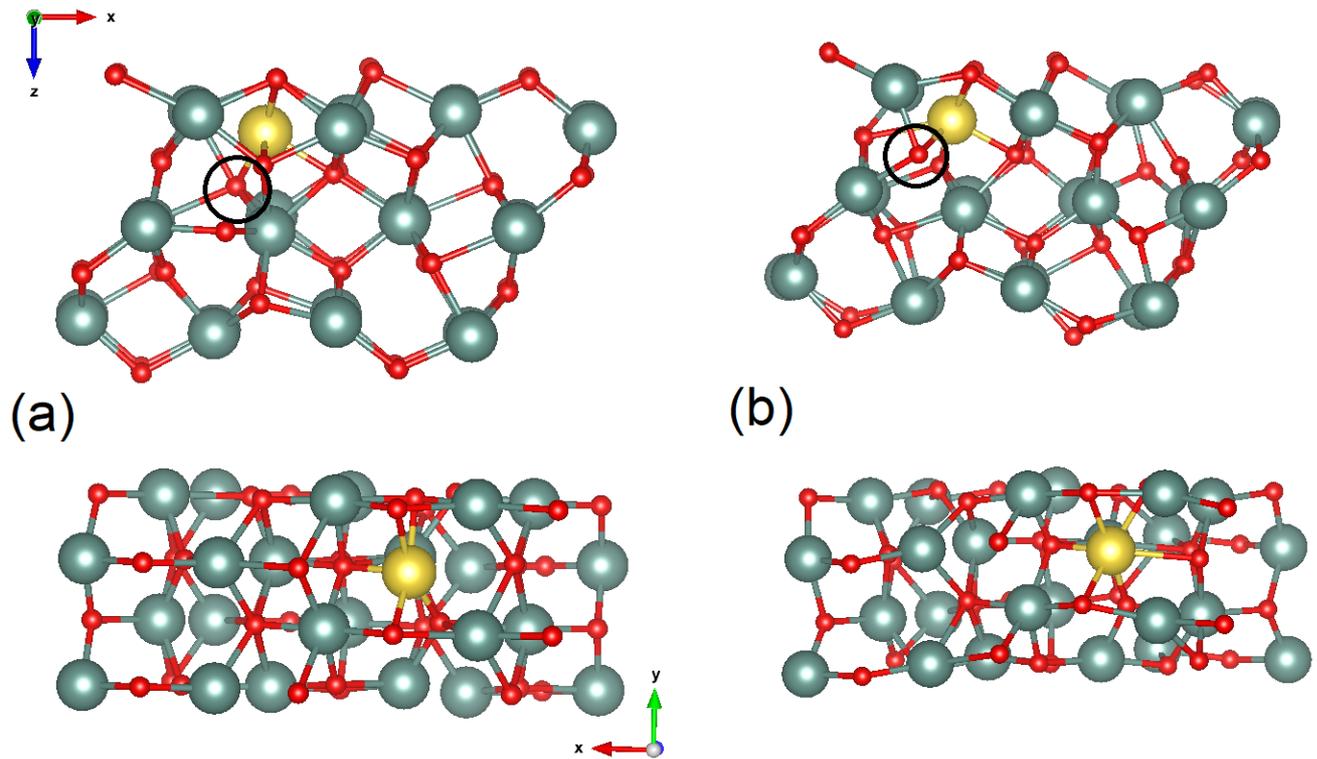

**Figure 7.** Optimized atomic structure of supercell employed for the modeling of bulk (a) and (010) surface (b) of monoclinic Y$_2$O$_3$ with interstitial sodium (large yellow spheres) and oxygen (shown as circles) defects. Yttrium atoms are shown by large grey-blue spheres; oxygen atoms are small red spheres.

Table II. Formation energies of various defects in the bulk Y$_2$O$_3$ for cubic and monoclinic phases. Also surface formation energies and oxygen vacancies (vO) in the surfaces of discussed systems are given. The most probable configurations of defects are shown in bold.

| Configuration | Phase | E$_{form}$ in bulk (eV/atom) | Surface (eV/Y2O3) | Surface+vO (eV/O) |
|---|---|---|---|---|
| defect free | cubic | . | 3.31 | -8.63 |
|  | monoclinic | . | 0.41 | 5.74 |
| vO | cubic | 6.68 | . | . |
|  | monoclinic | 6.67 | 0.39 | 6.05 |
| iO | cubic | 9.05 | . | . |
|  | monoclinic | **1.08** | 0.41 | **-1.03** |
| NaY | cubic | 11.16 | . | . |
|  | monoclinic | 10.67 | 0.37 | 2.02 |
| NaY+vO | cubic | 5.77 | . | . |
|  | monoclinic | 5.58 | 0.55 | 6.11 |
| NaY+iO | cubic | 9.67 | . | . |
|  | monoclinic | 5.40 | 1.89 | -0.63 |
| Nai | cubic | 12.89 | . | . |
|  | monoclinic | 4.70 | 0.59 | 5.35 |
| Nai+vO | cubic | 9.47 | . | . |
|  | monoclinic | 5.58 | 0.56 | 5.67 |
| Nai+iO | cubic | 5.74 | . | . |
|  | monoclinic | **2.04** | 0.55 | **2.08** |

Correct theoretical description of electronic structure and physical properties of thin layers requires simulation of both bulk and surface for studied materials. For this purpose we constructed the slab from supercells used for the simulation of defects in the bulk phase. Since all crystallographic surfaces are the same in cubic Y$_2$O$_3$, so for this reason we simulated (001) surface. In the case of monoclinic phase only the formation of (010) surface corresponds the absence of dangling bonds. The formation of defects in cubic Y$_2$O$_3$ is very unfavorable for this phase, thus we considered only the surface of defect-free slabs. For monoclinic phase, we performed simulations of the surfaces with all considered kinds of defects in bulk (see Fig. 7b for example). Also we performed simulations of oxygen vacancies formation in the surface layer. In order to evaluate such surface, we calculated the changes of total energies per formula unit of the systems as the bulk and the slab. On the one hand, the results of calculations performed (see Table II) demonstrate that formation of surface in cubic phase has relatively large energy cost, it is approximately 3.3 eV/Y$_2$O$_3$). From the other hand the formation of oxygen vacancy in surface layer is negative. Therefore, we can propose the formation of multiple oxygen defects in the surface of cubic $Y_2O_3$. Contrary, formation of (010) surface of monoclinic phase usually meets with rather moderate energy costs (within 0.5 eV/$Y_2O_3$) for almost all considered types of crystal structure. Formation energies of oxygen defects in these surfaces correspond to endothermic process for all cases except the presence of interstitial oxygen impurity. In the latter case the formation of oxygen defects can be discussed as removal of excessive oxygen from the lattice.

The last step of our study of the various impurities in $Y_2O_3$ is simulation of adsorption process of carbon mono and dioxide onto the surface of cubic and monoclinic phases (see Fig. 8). Calculated adsorption energies for CO are 0.33 and 0.59 eV/CO for cubic and monoclinic phases, respectively. As for $CO_2$, these values are 0.12 and 0.46 eV/$CO_2$ for cubic and monoclinic phases. Note, that decomposition of CO and $CO_2$ were not observed in the simulations performed (see Fig. 8). Hence experimentally observed contribution from carbon and CO bonds (see Fig. 5) can be attributed to robust physical adsorption of CO and $CO_2$ on the surfaces and grain boundaries of inspected $Y_2O_3$ layers.

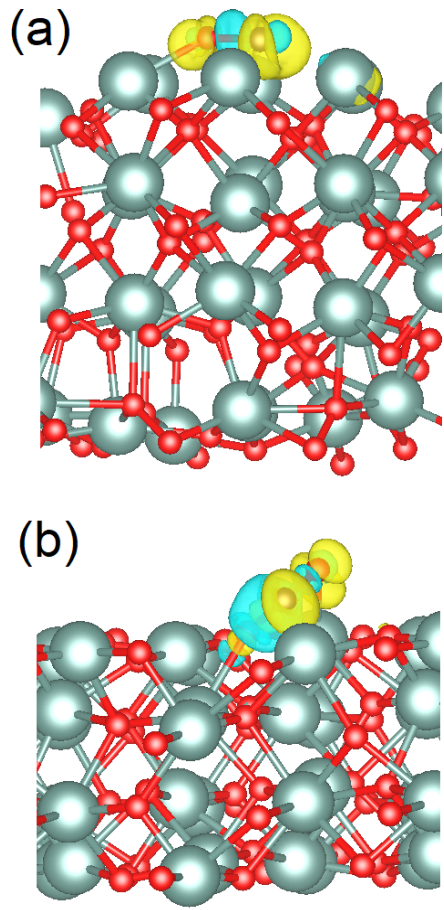

**Figure 8.** Optimized atomic structure and changes of charge densities for physical adsorption of CO on (001) surface of cubic (a) and (010) surface of monoclinic (b) lattices of $Y_2O_3$.

Yttrium atoms are shown employing large grey-blue spheres, whereas oxygen and carbon atoms are small red and brown spheres, respectively. Blue and yellow colors of charge density "clouds" correspond to the decreasing and increasing of charge density after adsorption.

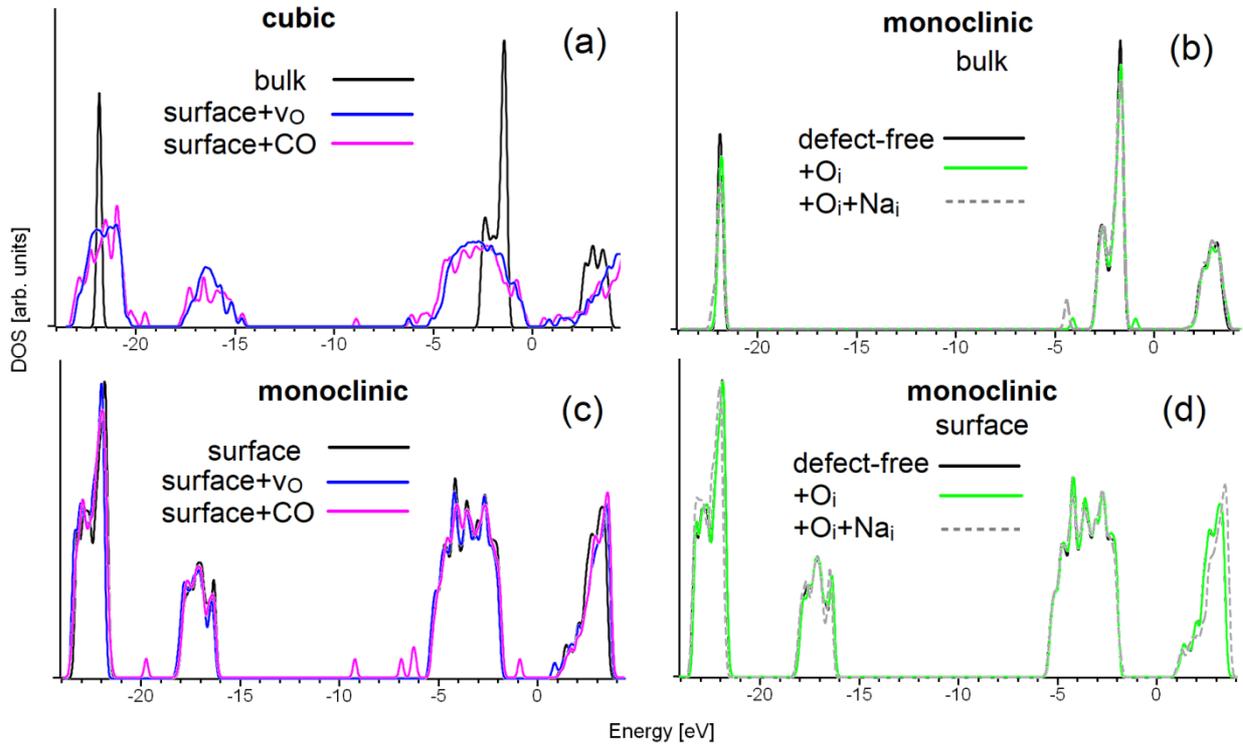

**Figure 9.** Total Densities of States (DOS) for the bulk morphologies (a,b) and surface (a,c,d) of $Y_2O_3$ with cubic (a) and monoclinic (b-d) lattices with and without various defects and surface impurities (see legends).

The last part of our theoretical studies of two phases of $Y_2O_3$ is the inspection of influence of surface and defects formation on the bulk surface of considered model systems. Results of the calculations performed demonstrate that electronic structure of bulk cubic and monoclinic $Y_2O_3$ is indistinguishable (see black lines on Fig. 9 a,b). Formation of the surface also affects similarly of both structures. The valence bands became broader with multiple peaks there and narrow in the bulk phase. At the same time Y 4p and O 2s bands also split and became broader (see Fig. 9 a,c). Note, that positions and widths of these peaks is the same for both considered phases. Two reasons of such changes in electronic structure can be proposed. The first one is appearance of surface and subsurface layers with electronic structure having differences from that for the bulk. The second is a violation of the symmetry in vicinity of surface (see Fig. 3 and Fig 7a in comparance with Fig. 7b and Fig. 8). Extremely narrow peaks in electronic structure of bulk

$Y_2O_3$ is an evidence of bands degeneration even in the presence of defects (see Fig. 9b). Thus we can propose the violation of symmetry in subsurface area as the leading reason of splitting energy levels of $Y_2O_3$ valence bands for both considered types of lattice.

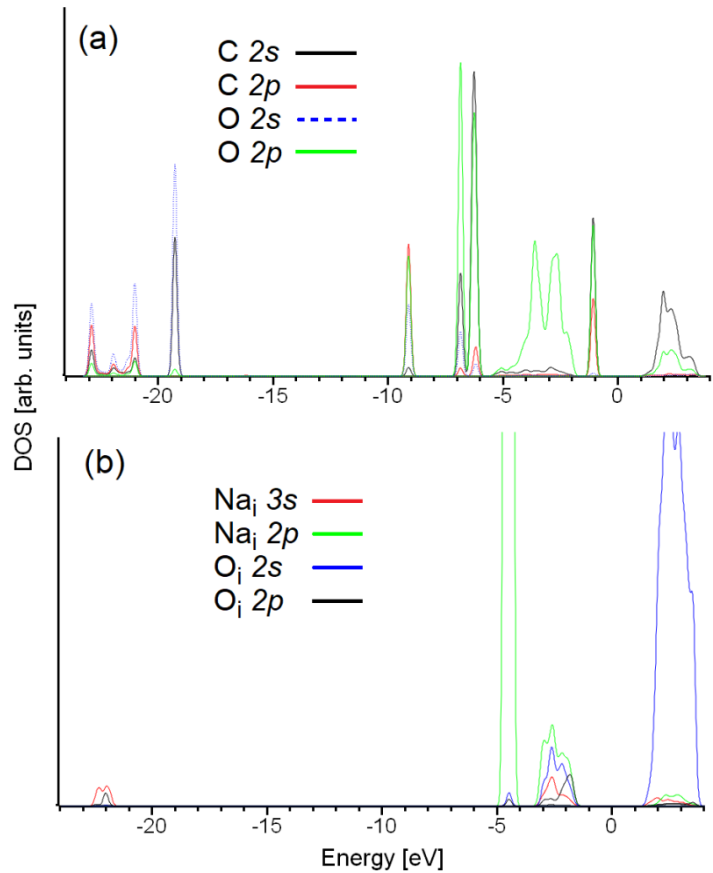

**Figure 10.** Partial densities of states for (a) carbon monoxide molecule adsorbed on (010) surface of monoclinic $Y_2O_3$ (see Fig. 8b) and interstitial sodium and oxygen impurities in bulk monoclinic $Y_2O_3$ (see Fig. 7a).

At the final step of our simulations we performed an inspection of the influence of adsorbed species and impurities on the electronic structure of yttrium oxide-based layers. Robust adsorption of carbon monoxide on the surface of $Y_2O_3$ provides appearance of similar small peaks in electronic structure for both considered phases (see magenta lines on Fig. 9a,c). These peaks can be clearly attributed with 2s and 2p states of carbon and oxygen of CO molecule (see Fig. 10a). Despite of relatively large magnitude of adsorption energies the influence of CO on valence band is miniscule. Redistribution of charge density after CO adsorption on the surfaces of $Y_2O_3$ (see Fig. 8) demonstrates significant changes in charge density of molecule which correspond to the appearance of some electronic states between -2 and -5 eV in electronic structure of CO (see Fig. 10a). Contrary changes of the charge densities in substrates are very local (Fig. 8). Thus we can conclude that adsorption of carbon mono and dioxides onto the surface and grain boundaries do not influence electronic structure of $Y_2O_3$ in both phases.

In the case of interstitial oxygen and sodium impurities in the bulk phase we also observe appearance of some minor peaks in electronic structure (see Fig. 9b). These peaks are the contribution from Na 2p and O 2p states (see Fig. 10b). Formation of the surface provides to the increasing of the number states near the edge of valence bands of $Y_2O_3$ (see Fig. 6) and peaks which are related with interstitial sodium and oxygen impurities disappear as separate features of the spectra.

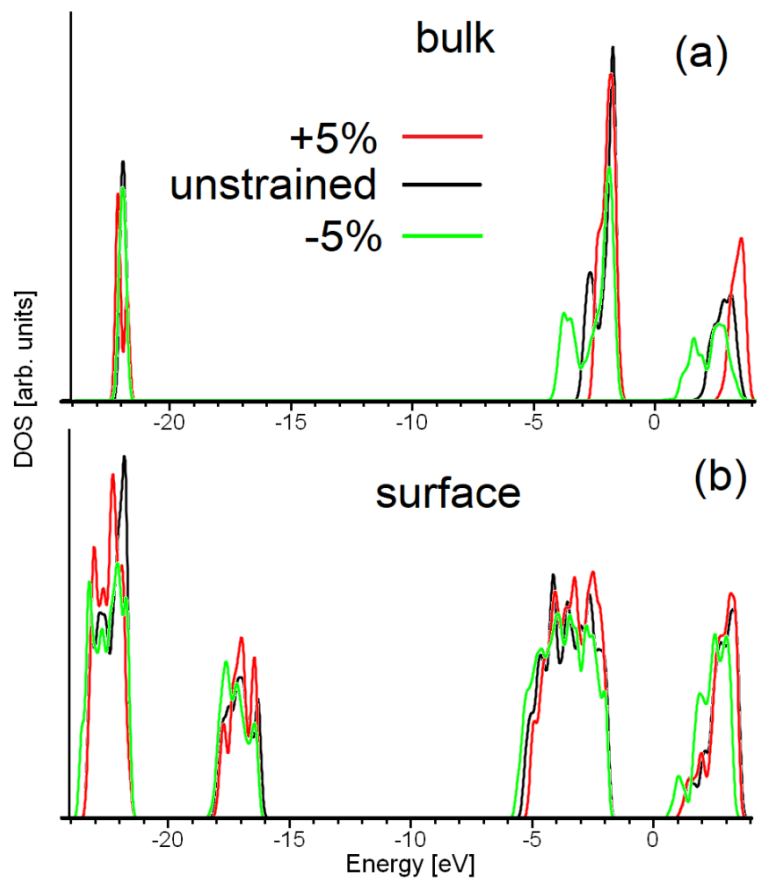

**Figure 11.** Total Densities of States (DOS) of unstrained (black lines), uniaxialy expanded by 5% (red lines) and uniaxialy compressed at 5% (green lines) bulk (a) and slab (b) of $Y_2O_3$ in monoclinic phase.

Since lattice distortion caused by the formation of surface provides larger impact on electronic structure of both studied phases of $Y_2O_3$, we also considered the influence of uniaxial expansion and compression on electronic structure of both bulk and surface $Y_2O_3$. As a case study we used monoclinic phase of $Y_2O_3$. In order to obtain more vivid effect we simulated rather large distortion of the lattice – at least 5%. Uniaxial expanding of $Y_2O_3$ lattice leads to an increase in the bandgap by approximately 1 eV and narrowing of the valence band (see Fig. 11a). Note that the position of VBM remains unchanged and narrowing of VB is provided by elimination of VB split of (see area around -3 eV in Fig. 11a). Contrary, uniaxial compression of lattice leads to a decrease of the bandgap by 1.2 eV and widening of VB by increasing of split between the peaks. Electronic structure transformations are less valuable in the case of uniaxial distortions of $Y_2O_3$ surface (Fig. 11b). In the case of uniaxial expanding only minor changes in the heights of the peaks can be observed. Compression the slab leads to similar minor redistribution of peaks heights and also results in decreasing of bandgap by value of 0.4 eV. Note that the magnitude of distortions used in these simulations is larger than that for lattice parameters caused by mismatch between lattice parameters of substrate and surface layers or some local distortions. Thus, we can omit strain effect on electronic structure and optical properties of realistic samples of $Y_2O_3$ for discussion. On the other hand, the simulations performed demonstrate crucial role of symmetry in $Y_2O_3$ electronic structure breaking process.

Summarizing theoretical part, we can conclude that electronic structure of bulk and surface $Y_2O_3$ is almost the same both for cubic and monoclinic phases. Neither stable adsorption of CO – $CO_2$ nor formation of defects provides visible changes in electronic structure of the surface for inspected phases. The main source of electronic structure variations for these structural phases of $Y_2O_3$ is established symmetry violation in vicinity of surface. The latter provides energy levels splitting and bands broadening.

### 3.3. Optical absorption and luminescence: bandgaps and excitonic states

Figure 12 shows optical absorption spectra of cubic and monoclinic $Y_2O_3$ layers in the Tauc coordinates for direct allowed interband transitions. The band gap was determined to be 5.76 eV and 6.12 eV for cubic and monoclinic $Y_2O_3$ films, respectively, using the Tauc equation $(\alpha \cdot h\nu)^n = A \cdot (h\nu - E_g)$, where $A$ is a constant, $\alpha$ denotes absorption coefficient, $h\nu$ is photon energy, $E_g$ is optical band gap and $n$ is the exponent depending on the type of transitions ($n = 2$ for direct allowed transitions [39]). It is known that for bulk cubic yttrium oxide the band gap is 6.0 eV [40]. Comparing the film and bulk forms of cubic yttrium oxide, it can be seen that the band gap for the film is somewhat smaller than for the bulk analog. We suppose that the cause of this discrepancy is due to the difference in the electronic structure of the bulk and surface $Y_2O_3$. As shown in Figure 9, in the electronic structure of $Y_2O_3$ surface, there is a splitting of levels and a broadening of the valence and conduction bands compared to the electronic structure of bulk $Y_2O_3$. This surface effect explains the observed narrowing of the band gap for cubic $Y_2O_3$ film compared to the bulk analog. Unfortunately, there is no information on the band gap for bulk monoclinic $Y_2O_3$ due to the technological complexity of preparing a single-phase sample with a monoclinic structure. However, based on the theoretical data shown in Section 3.2, we can assume the similar trend in the band gap difference between the bulk material and the layers for monoclinic yttrium oxide.

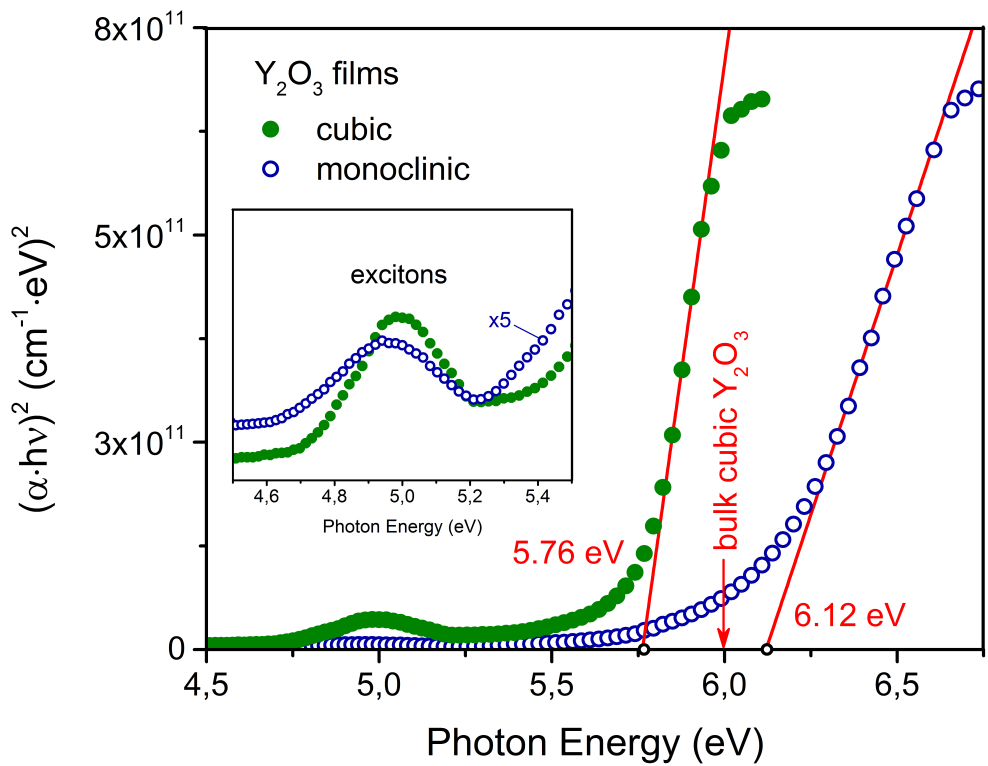

**Figure 12.** Optical absorption spectra of cubic and monoclinic $Y_2O_3$ layers in the Tauc coordinates for direct transitions. Red lines show approximation of linear ranges of spectra to determine the band gaps with an accuracy of ± 0.01 eV. Arrow indicates band gap (6.0 eV) for bulk cubic $Y_2O_3$. The inset shows an enlarged fragment of the spectra with band at 5.0 eV associated with exciton absorption.

Figure 13 shows normalized emission and excitation spectra of cubic and monoclinic $Y_2O_3$ layers. Excitation bands at 4.96 eV (for monoclinic $Y_2O_3$) and 5.00 eV (for cubic $Y_2O_3$) are associated with exciton absorption [23, 24]. It is known that the formation of exciton occurs due to absorbed photons with energy of 6.0 eV for the bulk cubic $Y_2O_3$ [41]. According to our findings, exciton absorption band for cubic $Y_2O_3$ layers is shifted to the low-energy region if compared with that for the bulk $Y_2O_3$. We believe that such a difference in spectral position of exciton absorption band for the layers and bulk forms of $Y_2O_3$ can be associated with differences in their electronic structure, as it was shown by theoretical modeling (namely, with the broadening of the valence and conduction bands, see Fig. 9).

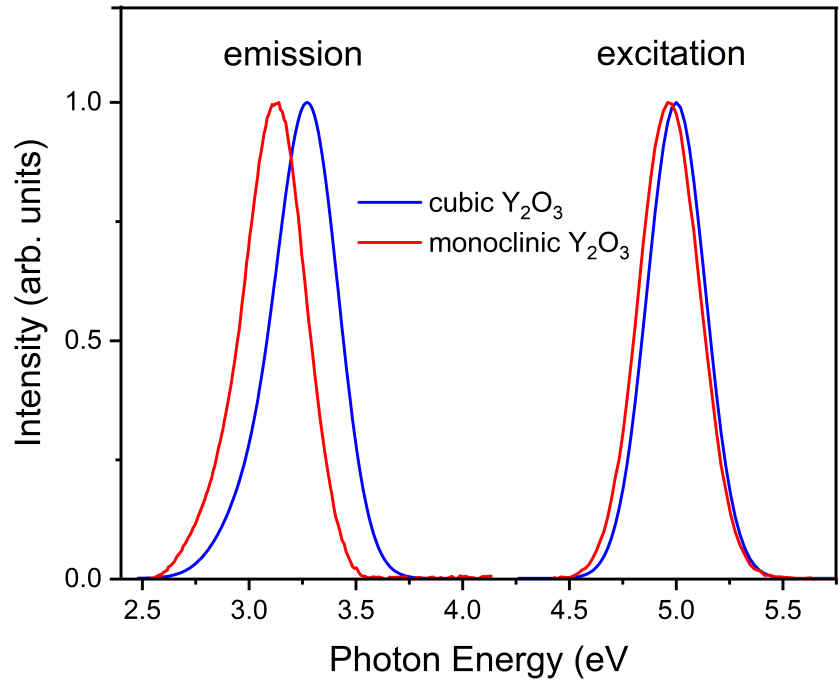

**Figure 13.** Excitons-related normalized spectra for cubic and monoclinic $Y_2O_3$ layers.

Emission spectra of both monoclinic and cubic $Y_2O_3$ layers are non-elementary and are represented by a superposition of two bands. The main contribution to emission spectra of monoclinic and cubic $Y_2O_3$ layers comes from the bands located at 3.13 eV and 3.27 eV, respectively (see Fig. 13). These emission bands are associated with radiative de-excitation of the self-trapped exciton [24, 42, 43]. Note also that for bulk $Y_2O_3$ the exciton emission band is shifted to the high-energy region (3.4-3.5 eV) [41]. The luminescence spectra of the layers also exhibit weak bands at 2.8 eV (for monoclinic phase) and 3.0 eV (for cubic phase). The origin of these bands was interpreted as radiative recombination of the bound exciton localized on the anion vacancy center [23, 24]. Comparing the spectra in Fig. 13, one can see that the spectral characteristics of the excitation-luminescence of layers with cubic and monoclinic structures are quite close to each other. This can be explained by the fact that electronic structure of $Y_2O_3$ is almost the same for cubic and monoclinic phase.

The obtained experimental data are in full agreement with results of theoretical modeling. The splitting of valence band energy levels observed in the electronic structure of surface $Y_2O_3$ with monoclinic and cubic structures is associated with the contribution of surface states. In our case the objects under study are nanocrystalline layers with high surface-to-volume ratio, therefore an additional contribution to the formation of surface states can also be made by grain boundaries. As shown by optical spectroscopy data, differences in the electronic structure of bulk and surface $Y_2O_3$ are reflected both in the band gap and in the luminescence characteristics. The revealed surface states can affect the ratio of the probabilities of radiative and nonradiative transitions in thin $Y_2O_3$ layers doped with rare earth activator ions.

## 4. Conclusions

Results of comprehensive experimental and theoretical studies of nano–crystalline (average size of grains about 10~14 nm) $Y_2O_3$ layers demonstrate insignificant difference in electronic structure and optical properties between two crystallographic phases (cubic and monoclinic). Note that the properties of monoclinic $Y_2O_3$ have not been studied before in details. Neither intrinsic (oxygen vacancies and interstitials) nor extrinsic (Na-impurities used for stabilization of monoclinic phase or carbon mono- and dioxide adsorbed on the surface) defects does not provide valuable contribution in electronic structure and optical properties. Contrary, local distortion and symmetry breaking caused by formation of the surfaces leads to splitting of the energy levels in valence and conductive bands that reduce the bandgap and increase the number of possible optical transition, which is corresponding with changes in emission and adsorption spectra. Form the other hand, the absence of valuable contribution from defects in optical spectra agrees with narrow adsorption and emission peaks.

Electronic structures and optical properties of both $Y_2O_3$ phases are almost identical, discussed in this work nanocrystalline layers of yttrium oxides could be grown on any chosen substrate without valuable variations of properties. Note that even huge strain (compression or expanding of the lattice at 5%) does not induce valuable changes in electronic structure of nanocrystalline $Y_2O_3$. The very high energy cost of intrinsic defects formation, negligible influence of intrinsic and extrinsic defects on electronic structure in combination with surface stability toward oxidation and carbon monoxide poisoning yields nanocrystalline $Y_2O_3$ films prospective material for light detection and conversion of electromagnetic energy.

**Acknowledgements**

The work has been supported by the Russian Science Foundation (project № 21-12-00392).